\documentstyle[preprint,aps]{revtex}

\begin{document}

\draft

\tighten

\preprint{\vbox{\hfill hep-th/9807138 \\
          \vbox{\vskip1.0in}
         }}

\title{IIB or not IIB}

\author{Mark Srednicki\footnote{E--mail: \tt mark@tpau.physics.ucsb.edu}}

\address{Department of Physics, University of California,
         Santa Barbara, CA 93106
         \\ \vskip0.5in}

\maketitle

\begin{abstract}
\normalsize{
We consider Type IIB superstring theory with the addition of $n$ 9-branes
and $n$ anti-9-branes (and no orientifolds).
The result is a ten-dimensional chiral theory of open and closed
oriented strings with gauge group $U(n)\times U(n)$.
There is, however, a tachyonic instability which can be understood
as the consequence of brane-antibrane annihilation.
We therefore expect to recover the usual IIB theory as the tachyon
rolls to infinity.
}
\end{abstract}


Many nonsupersymmetric string theories possess tachyons,
and there is a long history of speculation about their 
consistency (see \cite{banks} and references therein).
Of course, these theories are unstable against the tachyon rolling
over, but perhaps some other string theory is to be found at the bottom of
the potential.  As weak evidence in favor of this, 
we can note that all known tachyons have spin zero, 
and so can be represented as scalar fields at a local
maximum of the potential; a tachyon with nonzero spin would be more 
difficult (though perhaps not impossible \cite{ks}) to interpret.
Also, tachyonic theories have at least some of the duality properties
of their more stable cousins \cite{bg}.

D-branes (for a review, see \cite{tasi}) 
have provided a simple physical interpretation
of one class of tachyons: the instability is due to the annihilation
of coincident branes and antibranes.  Many examples of this
phenomenon have been constructed \cite{green,bs,gg,lif,gns,sen},
but so far all involve brane-antibrane annihilation as a local event
(in at least some of the dimensions).  
Here we investigate a fully ten-dimensional
version: annihilation of 9-branes and anti-9-branes.
Our original hope was that this would yield one of the known 
tachyonic theories \cite{dh,sw,aggmv}, but this turns out not to be the case.  
Instead, we find a previously unknown class of ten-dimensional
string theories in which the GSO projection in the open-string sector
is correlated with the Chan-Paton factor in such a way as to produce a 
$U(n)\times U(n)$ gauge symmetry.  This theory appears to be fully consistent,
except of course for the presence of the tachyon.  
Since the instability can be attributed
to the annihilation of 9-branes and anti-9-branes, we expect to find
the usual IIB theory as the tachyon rolls to infinity.
This scenario fits into the general framework of Sen \cite{sen}
for the restoration of supersymmetry via tachyon condensation.

We begin with weakly coupled IIB theory (which allows $p$-branes for odd $p$),
and add $n$ 9-branes and $n$ anti-9-branes, hereafter denoted as
$\overline 9$-branes.  We need equal numbers of each so that there
is no net RR 10-form charge.  (Recall that Type I theory
is constructed by adding 32 9-branes and one orientifold to IIB
theory, a configuration which also has zero net 10-form charge \cite{jp}.)
We expect the 9-branes to break 16 linear combinations 
of the 32 supersymmetries, specifically 
$Q_\alpha + \mu_9 {\widetilde{Q}}_{\dot\alpha}$,
where $\mu_9=+1$ is the normalized 10-form charge of a single 9-brane.
The $\overline 9$-branes should then break the remaining 16 supersymmetries.
The total energy density of the branes and antibranes is
\begin{equation}
{\cal E} = 2n\tau_9 = {2n \over (2\pi)^9 g \alpha'{}^5}\;,
\label{ev}
\end{equation}
where $\tau_9$ is the 9-brane tension
and $g \ll 1$ is the closed string coupling.
The ten-dimensional Newton constant is $G_{10} \sim g^2 \alpha'{}^4$,
and so the cosmological constant
$\Lambda = {G_{10}\cal E} \sim g/\alpha'$
is small in string units.

At tree level, the closed strings are undisturbed, so we have the
usual IIB spectrum.  The branes lead to four kinds of oriented 
open strings, which can be labeled by their endpoints as
9-9, 9-$\overline9$, $\overline9$-9, and $\overline9$-$\overline9$;
the corresponding Chan-Paton factors are
\begin{equation}
(\lambda^a)^i {}_j  \qquad
(\lambda^a)^i {}_{\overline\jmath}  \qquad
(\lambda^a)^{\overline\imath} {}_j  \qquad
(\lambda^a)^{\overline\imath} {}_{\overline\jmath} 
\label{chan}
\end{equation}
in an obvious notation.
We make the usual $(-1)^F = +1$ GSO projection
for the 9-9 and $\overline9$-$\overline9$ strings,
leading to massless bosons and fermions with $SO(8)$ chirality
$8_v + 8_s$ in the adjoint representation of $U(n)\times U(n)$.
We must then distinguish the 9-$\overline9$ and $\overline9$-9
strings in some way which involves a sign flip, and so we make the
opposite GSO projection, $(-1)^F=-1$, for these strings.  This
leads to a tachyon, and to massless fermions with $SO(8)$ chirality
$8_c$, in the $(n,\overline n) +  (\overline n,n)$ representation
of $U(n)\times U(n)$.

We can now check the consistency of this theory.  All purely gravitational
anomalies cancel, because we have added equal numbers of 
$8_s$'s and $8_c$'s to the usual IIB massless spectrum.
The nonabelian part of the pure gauge anomaly proportional to the single trace
$\mathop{\rm Tr}F^6$ also cancels, because
\begin{equation}
\mathop{\rm Tr_{\rm adj}}F^{2k} =  2n\mathop{{\rm Tr}_n}F^{2k} + \ldots
\label{tr}
\end{equation}
for $SU(n)$, where the ellipses stand for products of traces of smaller
powers of $F$, all with coefficients that are independent
of $n$ \cite{fk}.  This is just the relation we need.
The remaining anomalies all have a product of at least two traces
(or involve $U(1)$ factors), and so can in principle be
cancelled by the Green-Schwarz mechanism.
Note that there are three RR forms (of rank 0, 2, and 4) present,
all of which can participate in this cancellation.

Another possible problem is a sign ambiguity for amplitudes
involving a single gravitino vertex operator and an odd number
of open string vertex operators with the ``wrong'' GSO projection,
due to square-root branch cuts in the operator product expansion \cite{fms}.
However, the Chan-Paton factors for these open string states are
of the form 
$(\lambda^a)^i {}_{\overline\jmath}$ and 
$(\lambda^a)^{\overline\imath} {}_j$, and
so we must have an even number of them on any one worldsheet boundary.  
This eliminates the sign ambiguity.

Note that our theory has gravitinos coupled to matter fields
which are not supersymmetric; 
it is generally believed that such couplings are inconsistent.  
Here, though, we see a loophole: this theory is
properly viewed as an excited state of the standard IIB theory.
Excited states never have the full supersymmetry, and here there
is none at all, but the massless gravitinos remain.  

The tachyon potential (ignoring other fields, and setting
$\alpha'=1$ from here on) is of the general form
\begin{equation}
V(T) = -{1\over 2g} \sum_{n\ge1}c_n\,g^n \mathop{\rm Tr}(T^\dagger T)^n 
\label{v}
\end{equation}
where $c_1=+1$ and $c_2$ is known to be positive \cite{bs}.
As the tachyon rolls over, the gauge symmetry is spontaneously broken,
and the gauge bosons get a mass of order $g^{1/2}T$.
The gauginos, on the other hand, couple to the tachyon 
via a four-point amplitude, and get a mass of order $gT^2$. 
When $g^{1/2}T \sim 1$, the negative energy density of the tachyon
potential is $O(1/g)$, just like the original energy density of the
9-branes, eq.~(\ref{ev}).

Note that any expectation value for the tachyon leaves an unbroken 
diagonal subgroup of at least $U(1)^n$.  
The surviving subgroup  can in principle be further broken by the expectation
values of some other scalar fields, corresponding to massive spin-0
string states which are destabilized by the tachyon expectation value.
This can occur when $g^{1/2}T \sim 1$.  However, there is a $U(1)$ symmetry
which apparently remains unbroken:
the corresponding charge is the number
of 9-branes minus the number of $\overline 9$-branes, and all perturbative
string states are neutral under it.  (In this respect it is like
the $U(1)$ symmetry of the oriented open bosonic string.)
This symmetry can only be broken if some nonperturbative states 
(black holes?) carry it, and then condense as the tachyon rolls over.

Taking T-duals of this theory leads to various other configurations
of coincident branes and antibranes.  In particular, dualizing all
nine dimensions results in coincident 0-branes and anti-0-branes,
leading to a simple physical picture of this system.  

What lessons can we draw from this for other tachyonic theories?  
In the case at hand we have succeeded in adding extra states, 
including a tachyon, to an otherwise stable string theory.  
As the tachyon rolls over, these states should
all move up to infinite mass, and allow us to recover the original theory
(although as we have noted 
there is a $U(1)$ symmetry whose fate remains a puzzle).
If, on the other hand, other tachyonic theories are to turn into 
known superstring theories as their tachyons roll over,
then extra states (such as gravitinos) must
{\it appear\/} in the process, presumably by coming down from
infinite mass, and so precluding a chiral supergravity at the bottom.
Another alternative is that there are new, nonsupersymmetric,
stable theories which appear at the bottom of the tachyon potential.

Of course, the tachyonic theories could just be inconsistent after all,
but we hope the present example casts some doubt on this disappointing
possibility.

\vskip0.25in

{\em Note added:}  I have learned that the question
in the title of this paper was previously asked, in a different context,
in the title of \cite{grw}.

\begin{acknowledgments}

I am grateful to Joe Polchinski for many enlightening discussions,
and to Gary Horowitz for bringing ref.~\cite{grw} to my attention.
This work was supported in part by NSF Grant PHY--22022.

\end{acknowledgments}

\end{document}